\documentclass[%
 ,twocolumn%
 ,secnumarabic%
,amssymb, amsmath,nobibnotes, aps, prl]{revtex4}
\usepackage{docs}%
\usepackage{graphicx}
\usepackage{epsfig}
\usepackage{dcolumn}
\usepackage{bm}%

\expandafter\ifx\csname package@font\endcsname\relax\else
 \expandafter\expandafter
 \expandafter\usepackage
 \expandafter\expandafter
 \expandafter{\csname package@font\endcsname}%
\fi

\begin{document}
\title{\bf Surprising Sun}

\author{S. Turck-Chièze, $^{1}$ S. Couvidat, $^{2,1}$ L. Piau, $^{3}$ 
J. Ferguson, $^{4}$ P. Lambert, $^{1}$,  J. Ballot, $^{1}$ R. A. Garc\'\i a, $^{1}$ P. A. P. Nghiem, $^{1}$ }%
\affiliation{$^{1}$ SAp/DAPNIA/CEA, CE Saclay, 91191 Gif sur Yvette, France} 
\affiliation{$^{2}$ HEPL, 455 via Palou, Stanford University, CA 94305-4085, USA} 
\affiliation{$^{3}$ Institut d'Astronomie et d'Astrophysique, ULB, CP226, 1050 Brussels, Belgium} 

\affiliation{$^{4}$ Physics Department, Wichita State University, Wichita, KS 67260-0032, USA}

\date{\today}

\begin{abstract}
Important revisions of the solar model ingredients 
appear after 35 years of intense work which have led to an excellent agreement
 between solar models and 
solar neutrino detections. 
We first show that the updated CNO composition 
suppresses the anomalous position of the Sun
in the known galactic enrichment.
The following law: He/H= 0.075 + 44.6 O/H in fraction number is now compatible with
all the indicators. 
We then examine  the existing discrepancies 
between the standard model and solar - seismic and neutrino - observations  
and suggest some directions of investigation to solve them.
We update 
our predicted neutrino fluxes 
using the recent composition, new nuclear reaction rates and seismic models
as the most representative of the central plasma properties.
 We get $5.31 \pm 0.6 \, 10^6/cm^{2}/s$ 
for the total $\rm ^8B$ neutrinos, 66.5 SNU and 2.76 SNU for the gallium and chlorine detectors, all in remarquable agreement with
the detected values including neutrino oscillations for the last two.
We conclude that the acoustic modes and detected neutrinos 
see the same Sun, but that clear discrepancies in  solar modelling encourage further observational and theoretical 
efforts.  

\end{abstract}

\maketitle

The Sun is one of the best defined reference object in astrophysics.  As it is the most studied and best known star
in the Universe, the
main characteristics of the sun $-$ luminosity, mass, radius and composition $-$
are used as standard units in astrophysics. Through the years, progress has 
been constant to determine better
the different ingredients which enter in the description of a star: 
nuclear reaction rates, opacity coefficients, diffusion of elements, etc.
Two types of probes, (the solar acoustic modes and neutrino 
detections) have been particularly useful to check the internal properties of the Sun. 
The first probe determines the sound speed, the adiabatic exponent and the rotation profiles from which 
the amount of photospheric helium (due to the 
extraction of the adiabatic exponent), or the convective zone basis (due 
to the variation of the temperature gradient) are deduced. 

Precise acoustic modes have recently been used 
to predict neutrino fluxes through seismic models 
\cite{Turck2001, Couvidat2003}. 
In parallel, 
different flavours of neutrinos have been detected in the direction of 
the Sun in Sudbury Neutrino Observatory (SNO). So, the total neutrino flux associated to boron has been measured for the first time \cite{Ahmed2004}. 
These two improvements and their agreement have demonstrated the great 
interest to use the Sun as a laboratory for progressing on fundamental properties 
of the Universe. 

Nevertheless this satisfactory picture offers some contradictions. On one hand, it seems that the picture of the 
``standard'' model is a reasonable description of what we observe. It was noticed that 
the ``seismic models'' were not far from standard model. On the other hand, the 
Sun appears to be a more complex star than we thought for which one needs to interpret the internal rotation profile, 
the origin and evolution of the solar magnetic cycle(s) and the presence of meridional 
circulation.  It has so far been important to describe the thermodynamical status of the 
Sun; it is now a natural next step to reveal a dynamical picture of our star. 

In this paper, we discuss several aspects of the impact of the recent updates of the 
CNO 
composition abundaces (-20\% or -30\% depending of authors) and of the nuclear reaction rates 
for $\rm^7Be(p,\gamma)^8$B and 
$\rm ^{14}N(p,\gamma)^{15}O$ ( decrease by a factor 2)
on the knowledge of the Sun, following previous studies \cite{Turck2003, Bahcall2004, 
Basu2004}.
In Section 1, we show the status of the Sun  in the enrichment of stars in the galaxy. 
We present in Section 2 new models of the Sun, how they can be compared with seismic models
and possible 
interpretations and verifications of the discrepancies. Finally we recalculate 
neutrino predictions in Section 3 and show that we have a coherent picture of 
the Sun including the recent update.

\begin{figure*}
\vspace {1cm}
\includegraphics[width=8.5cm]{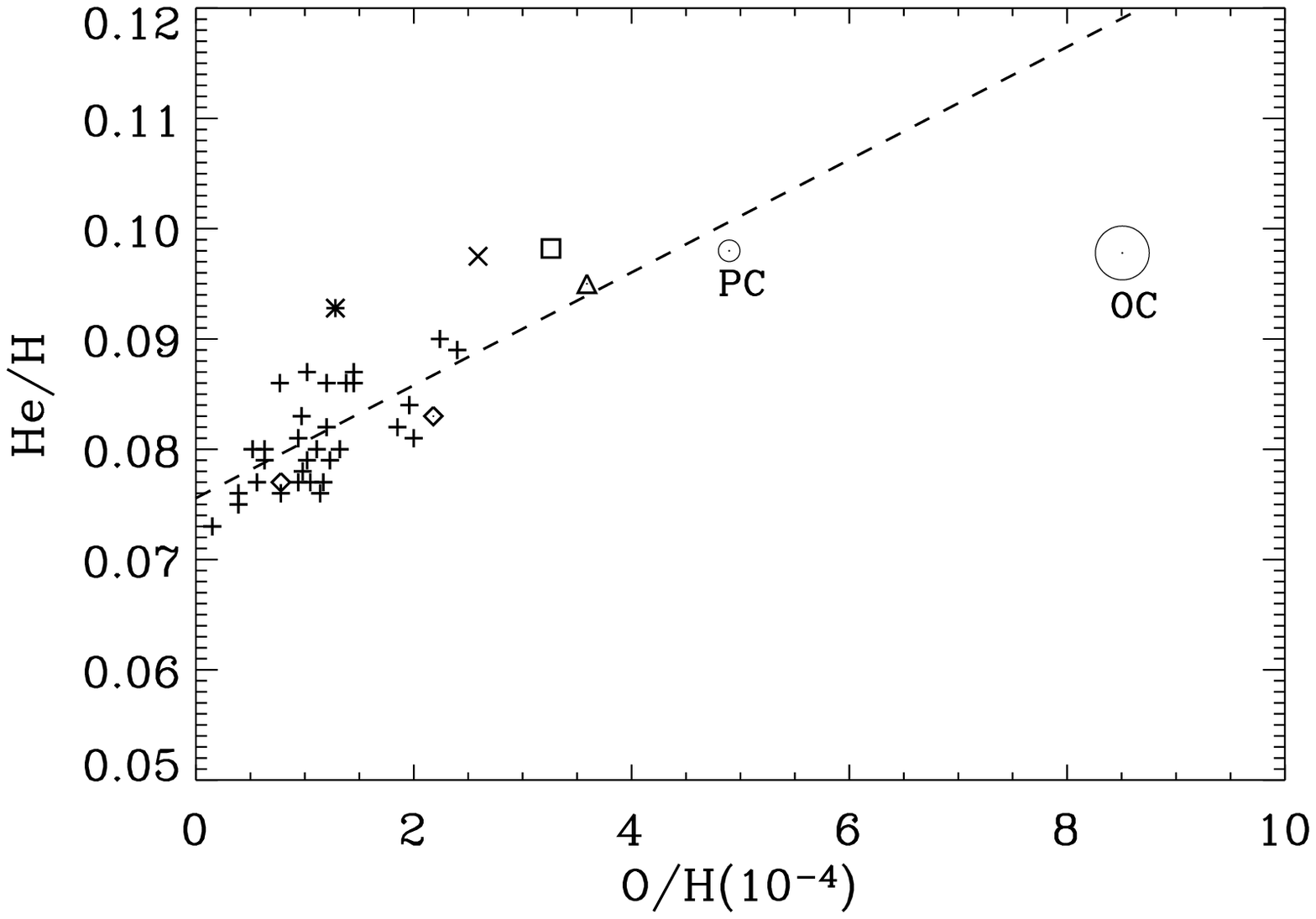}
\includegraphics[width=8.5cm]{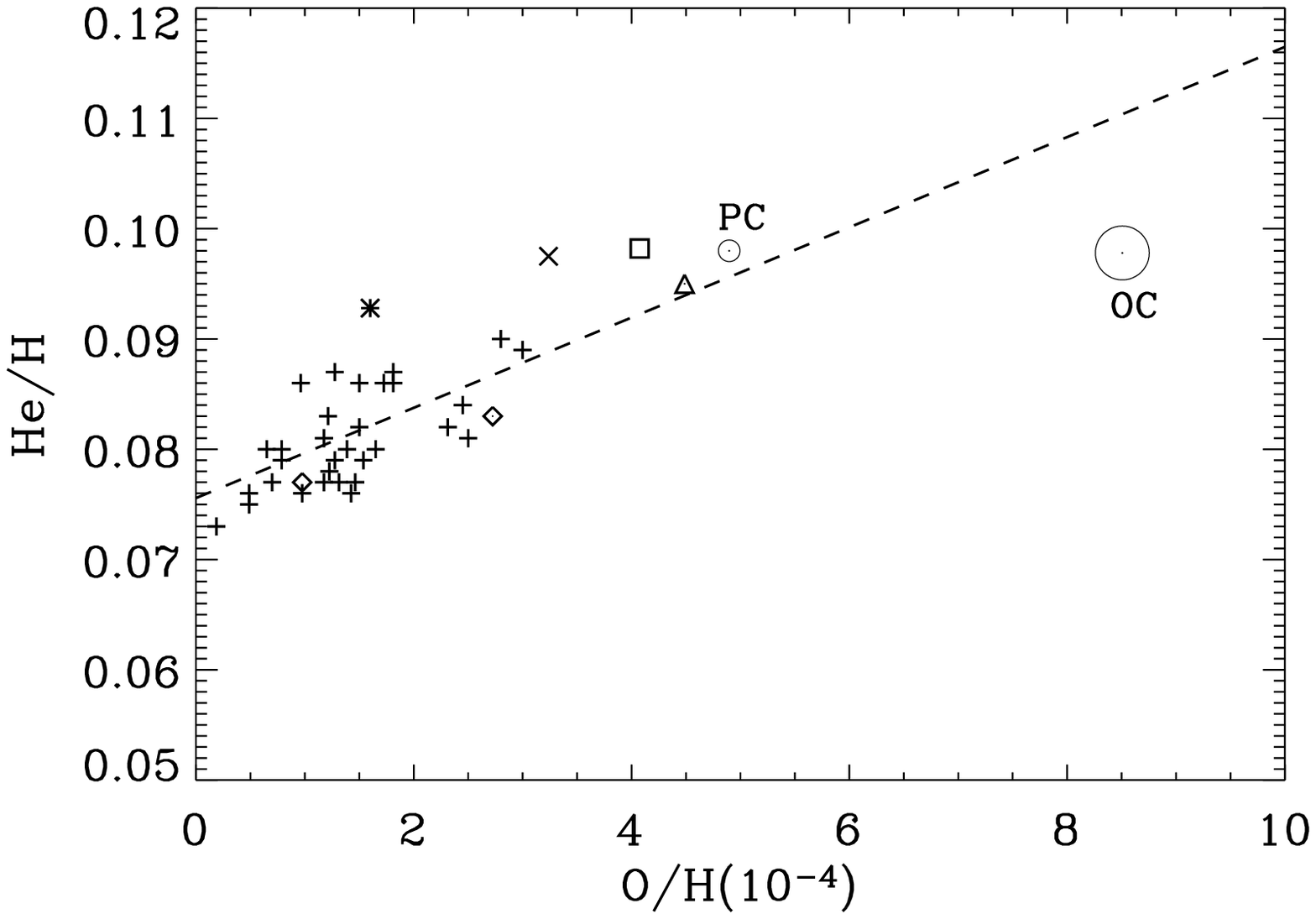}
\caption{\label{fig1}Evolution of the ratio Y/H (helium / hydrogen in fraction number) versus O/H for  
 extragalactic HII (+) , SMC (*), LMC (x,diamonds), M42 (square) and M17 (triangles) \cite{M17,M42,HII};
 the Present Composition [PC] for the Sun 
 is compared to the Old Composition [OC] \cite{Turck1993}. The left figure is extracted directly from observations, the secund includes a correction for oxygen 
 in grains recommended by 
 \cite{Meyer1989}}
\end{figure*}

\section{Galactic evolution and the Sun}

Fifteen years ago, the Sun appeared to be strangely rich in oxygen in comparison 
with its environment and with the Magellanic clouds \cite{Meyer1989, Turck1993}. Its metallicity was Z= 
0.02, where $Z$ is the mass fraction of elements heavier than helium, and the 
galactic enrichment in oxygen excluded the Sun 
as representative of the near neighborhood (Figure 1 [OC]). At that time, it has been
 suggested that the Sun
was formed in a cloud enriched by a supernova explosion. 
However, the comparison between meteoritic composition and 
photospheric composition \cite{Anders1989} revealed some contradictions. 

One of the contradictions has been solved by a reduction of 
the solar iron photospheric composition by 30\%  \cite{Grevesse1993}, so that the 
metallicity of the Sun has been slightly reduced (Z= 0.0173). As a consequence,
the central temperature has been reduced by 1.5\% due to the crucial 
role of iron in the opacity coefficient,  the $^8B$ neutrino flux has been reduced by 13\% and an increase of 
the discrepancy between model and the Sun for the sound speed profile has been noticed \cite{TurckLopes1993}. This effect has been compensated for 
by other progresses, for example the introduction of the microscopic diffusion (see the review of \cite{Turck2003}).
Today, CNO composition has been revisited and reduced by almost the same amount 
with a stronger impact on the metallicity (Z as low as 0.013). 
In the case of the oxygen, two origins to the overestimate of the abundance have been identified: a 
false contribution due to a previously unidentified nickel line  and the current use of hydrodynamical calculations of the
atmosphere which lead to a better coherence between the analysis of 
different lines \cite{Holweger2001,Asplund2004}.

 The solar initial helium abundance, obtained through a solar model, is not very sensitive
to the details of the models. So we can look to the impact of the recent oxygen measurement on the place of the Sun in the general
oxygen evolution along time (Figure 1). Contrary to the past situation, we note that the Sun appears now naturally enriched in 
oxygen in comparison with extragalactic HII regions, 
Magellanic clouds, other clusters and neighbours [PC]. 
 We can now deduce a galactic enrichment in oxygen, including the Sun, 
 after the introduction of  a correction for taking into account the 
oxygen locked in grains \cite{Meyer1989}. The best value we get is: 
 $ He/H= 44.6 \, O/H + 0.075$.
 
Recently, \cite{Gounelle2001} also noticed that the radioactive $\rm ^{26}Al$,  $\rm ^{10}Be$ and even $\rm ^7Be$ abundances 
in meteorites are compatible with production by irradiation in the disk of the young 
sun. They conclude that the presence of a supernova 
in the neighbourhood is not favoured. 

So these new estimates of the Sun composition solve serious problems
and must be taken as the result of 10 years of improvements in this field.

\section{Standard and seismic models}
The consequences of  the CNO abundance variations are well known: CNO play a role in the energy 
generation and consequently on the chlorine and gallium neutrino experimental 
predictions. They also play an important role in the opacity coefficients at all depths in the Sun 
but more specifically in the zone of the transition between 
radiation to convection, where the change 
in the degree of ionization of oxygen increases the opacity coefficient
(see \cite{Turck1993}). To test this impact,
solar models were computed with the 1D stellar evolution code CESAM using the most updated basic physical 
ingredients  already described in  \cite{Couvidat2003}. 
All the models are calibrated at the solar radius $R_{\odot}=6.9599 \times 
10^{10}$ cm, solar mass $M_{\odot}=1.9891 \times 10^{33}$ g, and solar luminosity 
$L_{\odot}=3.8460 \times 10^{33}$ erg, values at the age of $4.6$ Gyr 
including premainsequence. 
We also 
calibrate the photospheric metallicity, expressed by the ratio $Z/X$, where $X$ is the mass 
fraction of hydrogen; each model is calibrated at a specific $Z/X$ value. 

At 
low temperature ($T<5600$ K), we use opacity tables provided by coauthor, J. Ferguson, 
which were specifically calculated for this work and based on \cite{Ferguson}.
 These tables were computed for $Y=0.27$ (photospheric mass 
fraction of helium). For higher temperatures, we have computed three different 
sets opacity tables  from the OPAL website \cite{Iglesias}. 
The first opacity set is based on the abundances of Asplund (A) \cite{Asplund2004} 
for  C, N, O, Ne, and Ar elements, completed by the abundances of \cite{Grevesse1993}. 
The second set is based on the abundances of Holweger (H) \cite{Holweger2001} for 
 C, N, O, Ne, Mg, Si, and Fe elements, completed by the abundances of 
\cite{Grevesse1993}. A last set is based on the photospheric abundances of Lodders (L)
\cite{Lodders2003} Table 1, with  the isotopic abundances 
from Table 6, instead of the isotopic abundances of \cite{Anders1989}. 
Therefore we produce three kinds of solar models. 
For each kind we 
derive two models, one with mixing in the tachocline (transition region between radiation and convection,
prefix ``tac'' in the model 
name) and one without (prefix ``St'' in the model name).
The models are respectively calibrated at $Z/X=0.0172$, 0.0176, 0.0210 for 
Asplund, Lodders and Holweger composition and their main characteristics presented in Table 1, 
in comparison with seismic model results.

\begin{table}
\caption{\label{tab:fonts}Characteristics of the new models using  
Lodders, Asplund and Holweger compositions for standard (St) 
and model with turbulence in the tachocline (tac) compared with seismic model 2 of \cite{Couvidat2003}.}
\begin{ruledtabular}
\begin{tabular}{|l|c|c|c|c|c|}
 &  St A	     		&	tac L	&	tac H    &	tac A  & Seismic \\
\hline 
X$_{i}$ &	0.7195	 	   &	0.7245	&	0.7203	  &     0.7240 &  0.7064\\
Y$_{i}$  &	0.2664		&	0.2617	&	0.2633	  &     0.2625 &   0.2722\\
X$_{c}$ &	0.3526	 	&	0.3591	&	0.3522	&       0.3577 &  0.3371\\
Y$_{c}$  &	0.6323	 	& 0.6261	&	0.6301	  &     0.6278 & 0.6428\\
T$_{c}$  &15.58		&         15.495 &	           15.55	&  15.52 & 15.71\\
Y$_{s}$ &	0.2353	 	&	0.2400	&	0.2419	&      0.2407 & 0.251\\
$\alpha$	& 1.782	 	&	1.762	&	1.856	 &      1.754 & 2.04\\
BZC	& 0.7285		&	0.7307	&	0.7241	 &      0.7312  & 0.7113\\
(Z/X)$_s$	& 0.0172		 &   	0.0176	&	0.0210	 &      0.0172 & 0.0245\\
Ga	&120.9	  	&	118.3	&	121.6	&       119.0 &  126.8\\
  Cl	& 6.314	   	&	5.813	&	6.165	 &      5.956  &  6.9\\
Boron  &4.175  	&	3.801	&	3.982	  &     3.909  &  4.88\\
\end{tabular}
\end{ruledtabular}
\footnotetext[1]{ The indices i and s are for initial and surface, the central temperature Tc is in million degrees, 
boron flux in $\rm 10^6cm^{-2}s^{-1}$ }
\end{table}

\begin{figure}
\includegraphics[width=6cm]{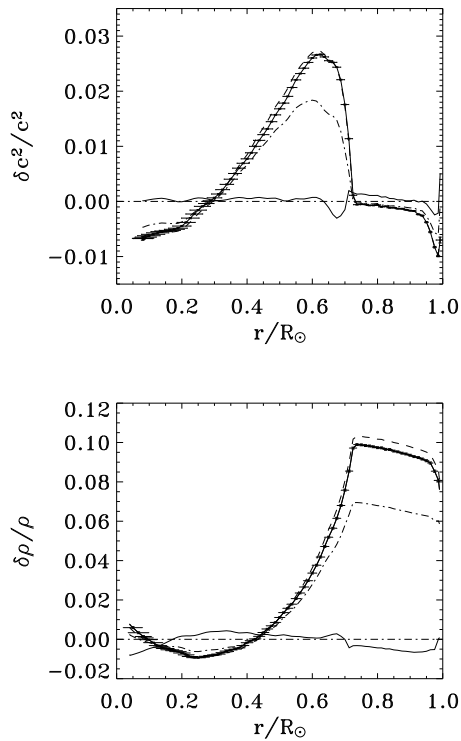}
\caption{\label{fig2}{Squared sound speed and density profile discrepancies 
between new updated standard models (full line with seismic error bar: 
tac A model, dot line: tac L model, dot dashed line: tac H model) 
and the seismic model (full line) with the present seismic observations\cite{Bertello}.}}
\end{figure}

We present in Figure 2 the models with turbulent mixing in the 
tachocline compared with the seismic model 2 of \cite{Couvidat2003}. 
This figure illustrates the discrepancies between these new models and the 
seismic results. As already mentioned \cite{Turck2003, Bahcall2004, Basu2004}, 
it is evident that the introduction of the 
new CNO composition largely deteriorates the previous agreement in the sound speed 
profile and does not improve the density profile in the 
radiative zone and particularly at the edge between the two types of energy transfer. 
In parallel the $\rm ^8B$ neutrino flux is largely reduced and is no more compatible with the SNO results. This 
does not mean that the new composition is incorrect, but that these models are 
not in agreement with the seismic observations. 

It could be partly due to the 
determination of the opacity coefficients in partially ionised elements.
It is interesting to note that Seaton and Badnell \cite{Seaton2004} show differences in their calculation 
in comparison with those of Livermore \cite{Iglesias}, which may explain part of the differences.  
Opacity coefficients are important ingredients of the solar model. So we recommend checking them with the new 
generation of high intensity lasers like the "Ligne Integration Laser" or future Laser MegaJoule or National Ignition Facility
as it has been done for lower temperatures and densities \cite{Chenais}. 
There is a clear need for experimental investigation in the 
million of degree range and density of fraction of g/cm$^3$. 

We cannot exclude  that  the introduction of the
microscopic diffusion is totally correct and one could need to improve it.
Another possibility is that the discrepancies are 
partly due to the absence of rotation effects in the radiative zone. 
Meridional circulation and magnetic field must be introduced
to justify a narrow sudden transition in the rotation profile 
\cite{Rudiger1997,McGregor1999}. 
Moreover a detailed energy balance must be looked for to check if 
 the nuclear energy balances precisely the surface luminosity. 
Is this new update of the composition the first evidence 
showing that the standard model is no longer representative 
of the present  Sun ?

\section{Revised neutrino predictions}
In our recent studies \cite{Turck2001, Couvidat2003} we have deduced 
 neutrino fluxes from the recent seismic results of SoHO \cite{Bertello}. 
 It is reasonable to think that 
these measurements are now sufficiently good in the region of emission of the neutrinos
to  give real insight into the properties and the mean central core temperature of the plasma.
  The seismic 
models we have built are not yet considered as physical models but they are representative models 
of the present seismic observations. They allow a determination of the main 
ingredients for neutrino predictions which are the temperature and density profiles 
in the radiative region.
 
At the same time, it is of great interest to improve the knowledge of the 
nuclear reaction rates as we improve the modelling of stars since such rate are essential 
ingredients necessary to
predict the neutrino fluxes.  Inversely it is also of great interest to 
extract the physical conditions of the core from the detected neutrino fluxes. So, 
recent revisions of the reaction rate $\rm ^7Be(p,\gamma)^8B$ or of 
$\rm ^{14}N(p,\gamma)^{15}O$, which is the slowest reaction of the CNO cycle, are extremely important.

$\rm ^7Be(p,\gamma)^8B$ has been remeasured several times these last few years, but without 
real agreement between measurements. This confirms the difficulty 
to determine this 
cross section, so a mean value between the recent measurements has been estimated by 
\cite{Junghans2003} of S(20 keV)= $20.7 \pm 0.8$ eV barn instead of the value 
of 19.4 eV barn \cite{Hammamache1998}
used in our previous predictions.
Using this revised value and seismic models, the new prediction for 
the $^8B$ neutrino is 5.31 $\rm 10^6 \pm 0.6 \,cm^{-2}s^{-1}$.
This value stays in complete agreement with the SNO result of 
$5.21 \pm 0.27 \pm 0.38$ \cite{Ahmed2004}, even if we need to introduce some magnetic field effect which has been 
considered as extremely small (2\%)
in our previous crude estimate \cite{Couvidat2003}. The uncertainties of this prediction have been slightly reduced with the 
recent progresses. The main contributor to the error is at present the knowledge of the $\rm (^3He,^4He)$ reaction rate which
 will be improved  with the upcoming LUNA planned experiment.

The new estimate of $\rm ^{14}N(p,\gamma)^{15}O$ astrophysical factor S(0) of $1.7 \pm 0.2$ keV b \cite{Formicola2004} 
instead of the recommended value  of 3.5 $^{+0.4}_{-1.6}$ \cite{Adelberger} is an important 
result for the lifetime of 
the hydrogen burning (increase by 0.7-1 Gyr of the globular cluster age). 
The CNO contribution to the luminosity decreases from 1.5\%
to  0.7\%, it is compensated by the pp luminosity, so the impact on the neutrino fluxes coming 
 from the pp chain is small. But this new estimate also influences
the neutrino predictions in the case of chlorine and gallium experiments. 
In fact, the $^{13}$N, $^{15}$O and $^{17}$F 
neutrino fluxes are doubly 
reduced by the effect of composition and reaction rate. 
They are reduced at  40\% of their previous values.

Consequently, we get $123.4 \pm 8.2$ SNU for the gallium prediction and $7.6 \pm 1.10$ SNU 
for the chlorine experiment. 
By applying the reduction on the neutrino fluxes due to LMA oscillation solution
$\Delta m^2 =7.3 \,10^{-5}$ and $tg^2 \theta _{12}$ =0.41 given by \cite{BahcallPeray}, 
we get respectively 66.65 SNU and 2.76 SNU (in solar neutrino unit ) 
for the detected fluxes which must be compared to $68.1 \pm 3.75$ SNU for combined gallium value 
\cite{CA} and $2.56 \pm 0.23$ SNU for the chlorine experiment. 

In conclusion current observations show a very good agreement 
between seismic predictions of neutrino fluxes and detected neutrinos but a difficulty 
to properly model the Sun. One needs to pursue the detailed observations of the radiative 
zone to orientate the main progresses to perform in modelling. 

We acknowledge financial support by NASA GRANTS NAG5-13261 and NAG5-12452 (SC) and by CNES for Saclay activities.

\end{document}